\documentstyle [12pt] {article}
\title{THERMODYNAMICAL CHARACTERISTICS OF "CRYSTAL LATTICE"
OF MANY INTERACTING KERR BLACK HOLES IN TOUCHING LIMIT}

\author{Vladan Pankovi\'c$^{\ast,\sharp}$,
Simo Ciganovi\'c$^\sharp$, Jovan Ivanovi\'c$^\sharp$\\
$^\ast$Department of Physics, Faculty of Sciences, 21000 Novi
Sad,\\ Trg Dositeja Obradovi\'ca 4. , Serbia, vdpan@neobee.net \\
$^\sharp$Gimnazija, 22320 Indjija, Trg Slobode 2a, Serbia \\}

\date {}
\begin{document}
\maketitle


\begin {abstract}
In this work, starting by simple, approximate (quasi-classical)
methods presented in our previous works, we reproduce effectively
and generalize final results of Herdeiro and Rebelo on the basic
thermodynamical characteristics (entropy and temperature) of two
interacting Kerr black holes (in touching limit) obtained recently
by accurate analysis. Like as it has been done in our previous
works, we simply suppose that circumference of the horizon of
total black hole (that includes two or, generally, a "crystal
lattice" of many interacting Kerr black holes in touching limit,
without angular momentum) holds integer number of reduced Compton
wave lengths corresponding to mass spectrum of a small quantum
system captured at horizon. (Obviously it is conceptually
analogous to Bohr quantization postulate interpreted by de Broglie
relation in Old, Bohr-Sommerfeld, quantum theory.) It, by simple
mathematical methods, first neighbour approximation of the black
holes interaction and first thermodynamical law, implies mentioned
basic thermodinamical characteristic of the total black hole as
well as any its part, i.e. single black hole. Especially, it is
shown that, in limit of increasing number of the black holes,
entropy and horizon surface of the total black hole stand
observables of the discrete spectrum while entropy and horizon
surface of the single black hole tends toward observables of the
continuous spectrum.
\end {abstract}
\vspace {1.5cm}

  Recently, Herdeiro and Rebelo [1] did an interesting, accurate analysis of two interacting Kerr black holes. In touching limit (corresponding to two horizons touching and when both, opposite angular momentums vanish) there are the following final results
\begin {equation}
   M_{1} = M_{2} = M
\end {equation}
where $M_{1}$ represents the mass of the first one and $M_{2}$
mass of the second one black hole,
\begin {equation}
   A_{1} = A_{2} = 2 (16 \pi M^{2}) = 2 (4\pi R^{2}) = 2A
\end {equation}
where $A_{1}$ represents the surface of the first one and $A_{2}$
surface of the second one black hole,
\begin {equation}
    S_{1} = S_{2} = 2 (4 \pi M^{2}) = 2 \pi R^{2}= 2\frac {A}{4} = 2S
\end {equation}
where $S_{1}$ represents the entropy of the first one and $S_{2}$
surface of the second one black hole,
\begin {equation}
    T_{1} = T_{1} = \frac {1}{2}\frac {1}{8 \pi M} =  \frac {1}{2} T
\end {equation}
where $ T_{1}$ represents temperature of the first one and $
T_{2}$ surface of the second one black hole, and where $M, R=2M, A
= 4\pi R^{2} = 16\pi M^{2}, S = \pi R^{2} = 4 \pi M^{2}$ and $T =
\frac {1}{8 \pi M}$ represent the mass, horizon radius, surface,
entropy and temperature of a Schwarzschild black hole.

In this work, starting by simple, approximate (quasi-classical)
methods presented in our previous works [2]-[5], we shall
reproduce effectively and generalize final results of Herdeiro and
Rebelo on the basic thermodynamical characteristics (entropy and
temperature) of two interacting Kerr black holes (in touching
limit). Like as it has been done in our previous works, we simply
suppose that circumference of the horizon of total black hole
(that includes two or, generally, a "crystal lattice" of many
interacting Kerr black holes in touching limit, without angular
momentum) holds integer number of reduced Compton wave lengths
corresponding to mass (energy) spectrum of a small quantum system
captured at horizon. (Obviously it is conceptually analogous to
Bohr quantization postulate interpreted by de Broglie relation in
Old, Bohr-Sommerfeld, quantum theory.) It, by simple mathematical
methods, first neighbour approximation of the black holes
interaction and first thermodynamical law, implies mentioned basic
thermodinamical characteristic of the total black hole as well as
any its part, i.e. single black hole. Especially, it will be shown
that, in limit of increasing number of the black holes, entropy
and horizon surface of the total black hole stand observables of
the discrete spectrum while entropy and horizon surface of the
single black hole tends toward observables of the continuous
spectrum.

Suppose that there is a series of $2N$ Kerr black holes
distributed along z-axis (representing a simple generalization of
two interacting Kerr black holes described by Herdeiro and Rebelo
[1]), where $N$ represents a natural number.

Suppose that initially, without touching limit, first black hole
holds the angular momentum $J_{1} = \frac {J}{2}$, second -
$J_{2} = -J$, third - $J_{3} = J$, …, $2N-1$-th - $J_{2N-1}= J$,
and $2N$-th - $J_{2N} = -\frac {J}{2}$.

Then, in the first neighbour approximation of the interaction
between black holes (and according to translation z-axis symmetry
characteristics for Herdeiro and Rebelo dynamics [1]), angular
momentum of any black hole slows down toward zero by interaction
with two closest neighbouring black holes. Simultaneously,
distance between any two neighbouring black holes tends toward
touching limit (when corresponding two horizons touching).

In touching limit, masses of individual black holes, $M_{n}$ for
$n = 1, 2, …, 2N-1, 2N$ , are mutually equivalent and equal
\begin {equation}
   M_{1} = M_{2} = … = M_{2N-1} = M_{2N} =M
\end {equation}
where $M$ represents the mass of a Schwarzschild black hole. Then
mass of total black hole, that includes all single black hole,
equals
\begin {equation}
    M_{tot}=2N M               .
\end {equation}

Suppose the following condition
\begin {equation}
    m_{tot \hspace{0.08cm} n} 2N (2 \pi R) = n
    \hspace{1cm} {\rm for}  \hspace{0.5cm} m_{tot \hspace{0.08cm} n} \ll M_{tot}  \hspace{0.5cm} {\rm and} \hspace{0.5cm}n=1,
      2,...
\end {equation}
where $ m_{tot \hspace{0.08cm} n}$ for  $ m_{tot \hspace{0.08cm}
n} \ll M_{tot}$ and    $n = 1, 2, …$  represent the masses, i.e.
mass (energy) spectrum of given small quantum system. It
corresponds to expression
\begin {equation}
      2N (2 \pi R) = n \frac {1}{m_{n}} = n \lambda_{tot \hspace{0.08cm} r \hspace{0.08cm} n}
      \hspace{1cm} {\rm for}  \hspace{0.5cm} m_{tot \hspace{0.08cm} n} \ll M_{tot}  \hspace{0.5cm} {\rm and} \hspace{0.5cm}n=1,
      2,...             .
\end {equation}
Here $2N (2 \pi R) = 2N (4\pi M) = 4\pi 2N M = 4\pi M_{tot}$
represents the length of a "N times 8"- like twisted but smooth
trajectory of given small quantum system at the total black hole
horizon equivalent to $2N$ times circumference of the single black
hole horizon, while
\begin {equation}
        \lambda_{tot \hspace{0.08cm} r \hspace{0.08cm} n}=\frac {1}{ m_{tot \hspace{0.08cm} n} }
\end {equation}
represents $n$-th reduced Compton wavelength of mentioned small
quantum system with mass $ m_{tot \hspace{0.08cm} n}$  for $n = 1,
2, …$ . Expression (8) simply means that {\it length of a "N times
8"- like twisted but smooth trajectory on the total black hole
horizon holds exactly} n {\it corresponding} n {\it-th reduced
Compton wave lengths of a small quantum system with mass} $ m_{tot
\hspace{0.08cm} n}$ {\it captured at the total black hole
horizon}, for $n = 1, 2, …$ . Obviously, it is essentially
analogous to well-known Bohr's angular momentum quantization
postulate interpreted via de Broglie relation. However, there is a
principal difference with respect to Bohr's atomic model. Namely,
in Bohr's atomic model different quantum numbers $n = 1, 2, …$ ,
correspond to different circular orbits (with circumferences
proportional to $n^{2} = 1^{2}, 2^{2}, …$). Here any quantum
number $n = 1, 2, …$ corresponds to the same "N times 8"-like
twisted but smooth orbit (with length $2N( 2\pi R)$ ).

According to (7) it follows
\begin {equation}
       m_{tot \hspace{0.08cm} n} = n \frac {1}{2N (2\pi R)} = n \frac {1}{8\pi M} = n \frac {1}{4\pi M_{tot}} \equiv n m_{tot 1}
       \hspace{1cm} {\rm for}  \hspace{0.5cm} m_{tot \hspace{0.08cm} n} \ll M_{tot}  \hspace{0.5cm} {\rm and} \hspace{0.5cm}n=1,
      2,...
\end {equation}
where $ m_{tot 1}$ represents the minimal small quantum system
mass
\begin {equation}
       m_{tot 1} = \frac {1}{2N (2\pi R)} =  \frac {1}{4\pi M_{tot}}        .
\end {equation}
Obviously, $ m_{tot 1}$ depends of $ M_{tot}$ so that $ m_{tot 1}$
decreases when $ M_{tot}$  increases and vice versa. For a
"macroscopic" black hole, i.e. for $ M_{tot}\gg 1$ it follows $
m_{tot 1}\ll 1 \ll M_{tot}$.

Now, we shall suppose that quotient of $ M_{tot}$ and $ m_{tot 1}$
represents Bekenstein-Hawking total black hole entropy, i.e.
\begin {equation}
       S_{tot} = \frac { M_{tot}}{ m_{tot 1}}= 4\pi M^{2}_{tot} = (2N)^{2} 4\pi M^{2}= (2N) ^{2} S          ,
\end {equation}
where, according to Bekenstein postulate,
\begin {equation}
       A_{tot} = 4 S_{tot} = 16 \pi M^{2}_{tot}= (2N) ^{2}  16\pi M^{2} = (2N) ^{2}  4S
\end {equation}
represents the total black hole surface area.

Differentiation of (12) yields
\begin {equation}
      dS_{tot} = 8\pi M_{tot}dM_{tot}
\end {equation}
or, after simple transformation,
\begin {equation}
      dM_{tot} = \frac {1}{8\pi M_{tot}} dS_{tot}                    .
\end {equation}
Expression (15), representing the first thermodynamical law,
implies that term
\begin {equation}
      T_{tot} = \frac {1}{8\pi M_{tot}} = \frac { m_{tot 1}}{2} = \frac {1}{2N}\frac {1}{8\pi M} = \frac {T}{2N}
\end {equation}
represents Hawking temperature of the total black hole, while $T =
\frac {1}{8\pi M}$, as it has been previously discussed,
represents Hawking temperature of a Schwarzschild black hole.

Further, since entropy represents an additive variable and since
all $2n$ single black holes are physically identical, it follows,
according to (29),
\begin {equation}
  S_{1} = S_{2} = … = S_{2N-1} = S_{2N}= \frac {S_{tot}}{2N} = (2N)  4\pi M^{2}= (2N)  S      .
\end {equation}

Finally, since according to Bekenstein postulate horizon surface
proportional to entropy represents an additive variable too, it
follows, according to (12),
\begin {equation}
  A_{1} = A_{2} = … = A_{2N-1} = A_{2N} = \frac {A_{tot}}{2N} = (2N)  16\pi M^{2}= (2N)  A   .
\end {equation}

It is not hard to see that our expressions (5), (17) and (18)
represent consequent generalization of Herdeiro and Rebelo
expressions (1), (3) and (2) respectively. Especially, for $N=1$,
i.e. $2N=2$, or for two interacting Kerr black holes in touching
limit, our expressions (5), (17) and (18) are identical to
Herdeiro and Rebelo expressions (1), (3) and (2) respectively.

Finally, it can be observed the following. By approximate changing
of the differentials by finite differences, (14) turns out
approximately, in
\begin {equation}
       \Delta S_{tot} = 8\pi M_{tot} \Delta M_{tot}
       \hspace{1cm} {\rm for}  \hspace{0.5cm} \Delta M_{tot} \ll M_{tot}  \hspace{0.5cm} {\rm and} \hspace{0.5cm}n=1,
      2,... .
\end {equation}

Now, assume
\begin {equation}
       \Delta M_{tot}= n m_{tot 1} \hspace{1cm}   {\rm for}  \hspace{1cm}  n=1,
    2,...
\end {equation}
which, according to (11), after substituting in (19), yields
\begin {equation}
       \Delta S_{tot} = 2n  \hspace{1cm}   {\rm for}  \hspace{1cm}  n=1,
    2,...
\end {equation}
or, according to (13),
\begin {equation}
       \Delta A_{tot} = (2n) 4 = (2n) 2^{2}  \hspace{1cm}   {\rm for}  \hspace{1cm}  n=1,
    2,...          .
\end {equation}
Obviously, expression (18) and (19) represent Bekenstein
quantization of the entropy and horizon surface of the total black
hole.

Since entropy and horizon surface represent additive variables it
follows that Bekenstein entropy and surface quantization for
single black hole, according to (21), (22), are
\begin {equation}
  \Delta S_{1} = \Delta_{2} = … = \Delta S_{2N-1} = \Delta S-{2N} =
  \frac {\Delta S_{tot}}{2N} = \frac {2n}{2N} \hspace{1cm}   {\rm for}  \hspace{1cm}  n=1,
    2,...
\end {equation}
\begin {equation}
  \Delta A_{1} = \Delta A_{2} = … = \Delta A_{2N-1} = \Delta A_{2N} =
  \frac {\Delta A_{tot}}{2N} = 4 \frac {2n}{2N}  \hspace{1cm}   {\rm for}  \hspace{1cm}  n=1,
    2,...
\end {equation}

It represents an interesting result. Namely, when number of the
black holes, $2N$, increases, entropy and horizon surface of the
total black hole, according to (12), (13), increase too, while
quants of entropy and horizon surface of the total black hole
stand the same and discrete according to Bekenstein quantization
rules (21), (22). In other words, in limit of increasing $2N$,
entropy and horizon surface of the total black hole stand
observables of the discrete spectrum. Simultaneously, when number
of the black holes, $2N$, increases, entropy and horizon surface
of the single black hole increase too according to (17), (18),
while quants of entropy and horizon surface of the single black
hole decrease according to (23), (24). In other words, in limit of
increasing $2N$, entropy and horizon surface of the total black
hole stand observables of the discrete spectrum while entropy and
horizon surface of the single black hole tends toward observables
of the continuous spectrum.

Thus, in conclusion it can be repeated and pointed out the
following. In this work we reproduce effectively and generalize
recent results of Herdeiro and Rebelo on the basic thermodynamical
characteristics (entropy and temperature) of two interacting Kerr
black holes (in touching limit) obtained recently by accurate
analysis. We started by simple supposition that circumference of
the horizon of total black hole (that includes two or, generally,
a "crystal lattice" of many interacting Kerr black holes in
touching limit) holds integer number of reduced Compton wave
lengths corresponding to mass (energy) spectrum of a small quantum
systems captured at horizon (that is conceptually analogous to
Bohr quantization postulate interpreted by de Broglie relation).
It, by simple mathematical methods, first neighbor approximation
of the black holes interaction and first thermodynamical law,
implies mentioned basic thermodynamic characteristic of the total
black hole as well as any its part, i.e. single black hole.
Finally, we showed that, in limit of increasing number of the
black holes, entropy and horizon surface of the total black hole
stand observables of the discrete spectrum while entropy and
horizon surface of the single black hole tends toward observables
of the continuous spectrum.

\section {References}

\begin {itemize}

\item [[1]] C. A. R. Herdeiro, C. Rebelo, {\it On the Interaction between Two Kerr Black Holes}, gr-qc/0808.3941
\item [[2]] V. Pankovic, M. Predojevic, P. Grujic, {\it A Bohr's Semiclassical Model of the Black Hole Thermodynamics}, gr-qc/0709.1812
\item [[3]]  V. Pankovic, J. Ivanovic, M. Predojevic, A.-M. Radakovic,
{\it The Simplest Determination of the Thermodynamical
Characteristics of Schwarzschild, Kerr and Reisner-Nordstr$\ddot
{o}$m black hole}, gr-qc/0803.0620
\item [[4]] V. Pankovic, S. Ciganovic, R. Glavatovic, {\it The Simplest Determination of the Thermodynamical Characteristics of Kerr-Newman Black Hole}, gr-qc/0804.2327
\item [[5]] V. Pankovic, R. Glavatovic, S. Ciganovic, D-H. Petkovic, L. Martinovic, {\it Single Horizon Black Hole "Laser" and a Solution of the Information Loss Paradox}, gr-qc/0807.1840

\end {itemize}

\end {document}